%% file: template.tex
\documentclass{article}
\pdfoutput=1

\usepackage{arxiv}

\usepackage[utf8]{inputenc} % allow utf-8 input
\usepackage[T1]{fontenc}    % use 8-bit T1 fonts
\usepackage{url}            % simple URL typesetting
\usepackage{booktabs}       % professional-quality tables
\usepackage{amsfonts}       % blackboard math symbols
\usepackage{nicefrac}       % compact symbols for 1/2, etc.
\usepackage{microtype}      % microtypography
\usepackage{lipsum}		% Can be removed after putting your text 
\usepackage{amsmath}
\DeclareMathAlphabet{\mathpzc}{OT1}{pzc}{m}{it}

\usepackage{mathrsfs} % for \mathscr{T}
\usepackage{graphicx}
\usepackage{hyperref}       % hyperlinks

\usepackage[caption=false]{subfig}

\usepackage{epstopdf}
\title{Random Projections of Mel-Spectrograms as Low-Level Features for Automatic Music Genre Classification}

\author{
  Juliano H. Foleiss \thanks{Also as a PhD candidate at the School of Electrical and Computer Engineering, University of Campinas -- Brazil.}\\
  Department of Computing\\
  Federal University of Technology -- Paraná\\
  Campo Mourão, PR -- Brazil \\
  \texttt{julianofoleiss@utfpr.edu.br} \\
   \And
  Tiago F. Tavares \\
  School of Electrical and Computer Engineering\\
  University of Campinas\\
  Campinas, SP -- Brazil \\
  \texttt{tavares@dca.fee.unicamp.br} \\
}

\begin{document}
\maketitle

\begin{abstract}
In this work, we analyse the random projections of Mel-spectrograms as low-level features for music genre classification. This approach was compared to handcrafted features, features learned using an auto-encoder and features obtained from a transfer learning setting. Tests in five different well-known, publicly available datasets show that random projections leads to results comparable to learned features and outperforms features obtained via transfer learning in a shallow learning scenario. Random projections do not require using extensive specialist knowledge and, simultaneously, requires less computational power for training than other projection-based low-level features. Therefore, they can be are a viable choice for usage in shallow learning content-based music genre classification.
\end{abstract}

% keywords can be removed
\keywords{Random Projections \and Music Genre Classification }

\input{randomproj}

\bibliographystyle{unsrt}  
\bibliography{tese}  %%% Remove comment to use the external .bib file (using bibtex).

\end{document}

%% file: randomproj.tex
\section{Introduction}
Automatic music genre classification (AMGC) is the task of labelling music tracks according to their genre (e.g., rock, reggae, and classical) \cite{tzan2002}. 
Audio-based music genre classifiers have received many contributions in the last two decades, including the development of new texture descriptors \cite{tao2003,peeters2004,lidy2005,manaris2008} and novel neural network architectures \cite{choi2017,yandre2017,yu2019}. They all rely on the assumption that each music genre can be characterised by their use of specific techniques and instruments, which lead to genre-specific sets of auditory textures \cite{tzan2002}. For this reason, most music genre classifiers rely on mapping each musical track from a collection to a point within a feature-based vector space whose topology can represent the perceptual similarities between tracks.

Track similarity was modelled in earlier work on AMGC using features inspired on domain-knowledge relevant concepts. Such handcrafted feature sets have been used as basis for a wide range of automatic music genre classifiers \cite{tzan2002,tao2003,peeters2004,lidy2005,manaris2008,shin19,kobayashi18}. However, their development involves considerable effort on modelling perceptual or musical characteristics of audio signals.

This drawback has been approached in the last few years by contributions towards feature learning using deep neural networks \cite{nanni2016,choi2016,yandre2017,oramas17,choi17b}. These methods rely on learning features that optimally correlate to the target labels. This process depends on a high amount of computational power for processing, which leads to higher hardware requirements.

In this paper, we evaluate a different paradigm for generating low-level features, namely the use of random projections of Mel-spectrograms. For such, we use a simple classification pipeline that begins with low-level feature extraction, proceeds to feature aggregation and ends with a vector classification. This pipeline was executed over five publicly available datasets, showing that, in commercial music datasets, random projection results in performance comparable to learned features, and outperform both handcrafted features and features obtained by transfer learning techniques.

These results indicate that random projections can be used in media organisation problems related to user customisation. In these problems, it can be unfeasible to execute feature learning techniques due to processing power and energy constrains. These problems can be overcome by using random projections, leading to consistent results while requiring significantly lower processing power.

Random projections have been used in previous work on automatic genre classification, specifically in Extreme Learning Machines (ELMs) during the vector classification stage \cite{Scardapane2014,loh2006,baniya2015}. Also, work by Chang et al. \cite{chang2010} has proposed using random projections of handcrafted features as basis for automatic music genre classification. Chang et al.'s \cite{chang2010} work was later commented by Sturm \cite{sturm2013}, who showed that the random projections do not add classification power to the handcrafted features themselves.

A similar idea, studied by Choi \cite{choi17b}, is to set the weights of a convolutional neural network to random numbers. This allowed highlighting a small performance gain obtained by learning features from other domains.

This work differs from both ELMs \cite{Scardapane2014,loh2006,baniya2015} and Chang et al.'s \cite{chang2010} work because it proposes using random projections to generate the low-level features that are later fed to a vector classification pipeline. Also, differently from Choi's \cite{choi17b} work, it is not bounded to a specific classification algorithm, thus it can be immediately applied in future research.

The results obtained in this work show that the low-level random projection features lead to better classification results than the original Mel-Spectrogram. This indicates that this usage of random projections is not harmed by the effects detected by Sturm \cite{sturm2013}.

Finally, we evaluated using learned features in a cross-dataset scenario. In these tests, the results were highly degraded, and the proposed random projections outperformed the learned features.

The remainder of this paper is organised as follows. Section \ref{sec:rp_random} presents a brief overview of the random projection theory applied to classification. Section \ref{sec:rp_method} presents our approach to evaluate random projections of Mel-Spectrograms as features for AMGC. Section \ref{sec:rp_datasets} presents the datasets used in the experiments, along with some remarks regarding fold creation procedures. Section \ref{sec:rp_results} presents results and discussions on the performance of random features in five different datasets and how they compare to other feature sets. Finally, Section \ref{sec:rp_conclusion} presents some closing remarks.

\section{Random Projections} 
\label{sec:rp_random}

\begin{figure}[h]
    \centering
    \includegraphics[scale=.95]{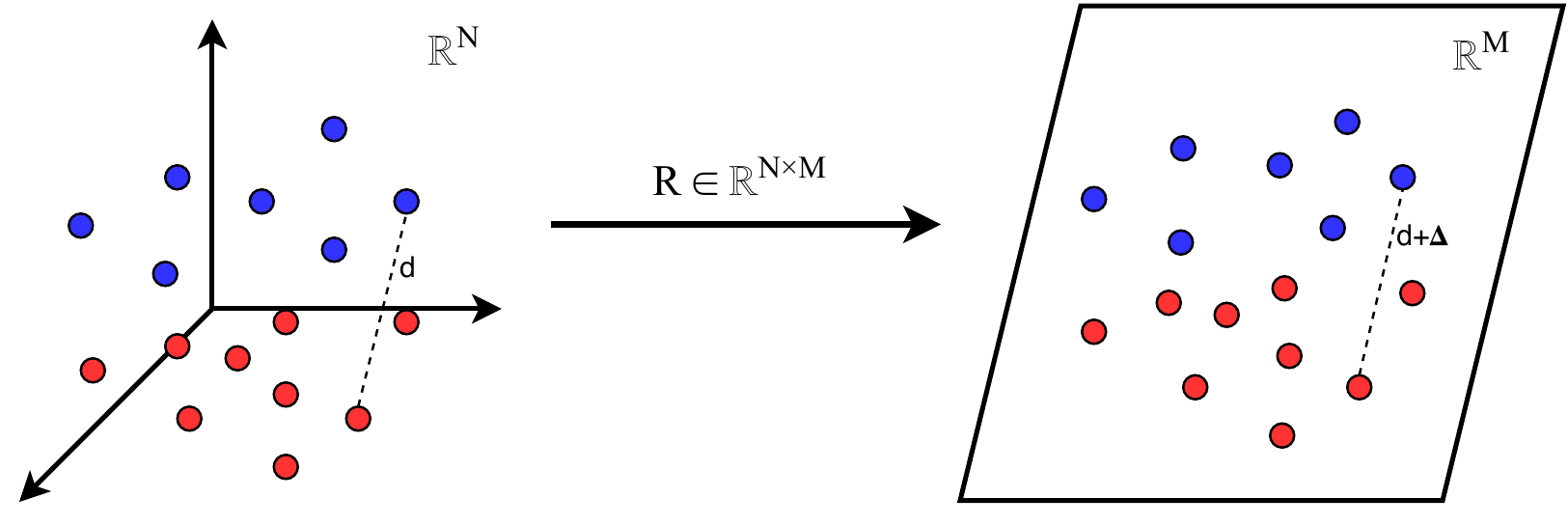}
    \caption{Random Projection. A cloud of points in $R^N$ are transformed into points in $R^M$ by a random linear transformation, while maintaining distance topology. This is known as a \textit{stable embedding}. (Adapted from \cite{baraniuk2010})}
    \label{fig:randomproj_rp}
\end{figure}

The effectiveness of random projections for dimensionality reduction is known well-known in the machine learning literature \cite{baraniuk2010}. The Johnson-Lindenstrauss (JL) lemma \cite{johnson1984} states that a random matrix $R \in \mathbb{R}^{N \times M}$, where $N > M$, projects a matrix $A \in \mathbb{R}^{T\times N}$ into a stable embedding $A \in \mathbb{R}^{T\times M}$ with high probability if $M = O(\log(N) \epsilon ^{-2})$, $\epsilon \in (0,1)$ \cite{johnson1984}. Figure \ref{fig:randomproj_rp} shows this transformation. In a machine learning scenario, $T$ is analogous to the number of feature vectors, $N$ is the original dimensionality, and $M$ is the dimensionality of the embedding. The $\epsilon$ constant controls the distortion introduced by the random transformation. It follows that, as more distortion (higher $\epsilon$) is allowed, the smaller $M$ can be.

When data is projected into a \emph{stable embedding}, the distances among the points in the embedding are preserved in relation to the data in the original domain \cite{baraniuk2010}. Figure \ref{fig:randomproj_rp} shows how the relative distance among all points are preserved. This property enables vector-based classification to be performed in the embedding domain \cite{davenport2007}. In addition, the dimension reduction caused by the projection (when $M < N$) reduces the computational effort and alleviates the curse of dimensionality.

Baraniuk \emph{et al.} \cite{Baraniuk2008} have shown a link between the JL lemma and the Restricted Isometry Property (RIP) of compressive sensing theory. They state that random matrices satisfy the RIP property as a consequence of the JL lemma. Also, the RIP property guarantees that a \emph{sparse signal} can be recovered from its sampled form using less components than the lower bound introduced by the JL lemma \cite{Calderbank2009,candes2008}

Mel-Spectrograms can be considered sparse, because many components of each frame are close to zero. Hence, Mel-Spectrograms are fit for compression into a lower-dimensional space using random projections. Also, since the final objective is classification, instead of signal reconstruction, it can be expected \cite{lohit2015} to be possible to use fewer dimensions in the embedding domain ($M$) than the lower bound presented by Candes and Wakin \cite{candes2008}. 

\section{Proposed Method}
\label{sec:rp_method}
The classification method used in our work consists on mapping each track to a vector representation. This representation aims at preserving auditory similarities, that is, tracks that sound similar should be mapped to vectors that are close to each other in feature space. For such, we used four different low-level feature sets.

All feature sets were calculated using a $23$ ms STFT with a Hanning window and $50$\% overlap between subsequent windows. The first and second order differentials of each feature were also calculated. Then, each feature is aggregated to texture-level frames using mean and variance calculated over a $2.5$s-long sliding window. Last, the mean and variance of the mean and variances are used as features for classification.

The first feature set used comprised subset of MARSYAS \cite{tzan2002}, which included Energy, Spectral Centroid, Spectral Rolloff, Spectral Flatness, Spectral Flux, Zero-Crossing Rate, and the first 20 MFCC coefficients. Theses feature were handcrafted based on domain-specific knowledge.

We also evaluated features learned using an auto-encoder over the Mel-scale spectrogram (MEL-AE). For such, we used the activations from the bottleneck layer as features. The architecture consisted of an input layer with one input for each MEL-SPEC bin, followed by a fully-connected layer with $H$ units using ReLU as the activation function. The final layer is a fully connected layer using a linear activation function containing a unit for each MEL-SPEC bin. The Auto-Encoder network was trained using \emph{Nesterov momentum} with learning rate $0.01$ and momentum $0.9$, $40000$ samples per batch for a maximum of $200$ epochs, using an early stopping criterion of $50$ epochs with no improvement. $H$ is a parameter which we tested using $H \in \{ 16, 32, 64, 128, 256\}$ units.

This procedure was also used to generate features used in transfer learning settings. In these settings, we trained the auto-encoder using a dataset and then performed classification experiments using a different dataset. The training and testing procedures comprised the GTZAN and LMD datasets, because they contain tracks with non-overlapping labels.

Our proposal, MEL-RP, consists of using a random projection of the Mel-scale spectrogram as a feature set. The projection matrix was drawn element-wise from a Gaussian distribution with zero-mean, unit-variance. The target dimensionality $M$ was tested for $M \in \{8, 26, 51, 75, 100\}$.

For baseline purposes, we used the 128-bin Mel-Spectrogram  (MEL-SPEC) itself as a frame-level feature set. Last, we used a PCA-based feature reduction of the Mel-Spectrogram (MEL-PCA). The target dimensionality for the PCA-based feature reduction was $\{8, 26, 51, 75, 100\}$, the same set as the experiments done with the random projection.

For classification, we tested a Support-Vector Machine (SVM) and a K-Nearest Neighbors (KNN) classifier. Their hyper-parameters were adjusted with a 80-20 train/validation scheme in the training set. The SVM used a RBF kernel, C was optimised over $\{1, 10, 1000, 10000\}$, and  gamma was set to $1/\text{(\# features)}$. The KNN had its K parameter optimised over $\{1, 5, 10, 20\}$. After the hyper-parameter estimation, the whole training set is used to train the highest performing model for each classifier.

Before training the classifier, the training set is normalised so that all features are centred at zero mean and unit variance. Test samples were normalised using the same parameters used on the training set.

\section{Datasets}
\label{sec:rp_datasets}
The datasets used in the experiments are shown in Table \ref{tab:rp_datasets}. All datasets were resampled to 44100Hz and mixed into monaural tracks by averaging the stereo signals. The experiments were conducted using specific train-test splits for each dataset, allowing comparison with previous work using the same datasets. Since many datasets have repeated songs from the same artist, we applied an artist filter \cite{pampalk2005} when creating the cross-validation splits for every dataset to prevent modelling artist-specific (instead of genre-specific) characteristics.

\begin{table}[htbp]
\centering
\caption{Dataset description and corresponding train-test split procedures used in the experiments, including the number of folds for stratified cross-validation.}
\begin{tabular}{@{}llllll@{}}
\toprule
Dataset & Tracks & Classes & Balance & Clip Len. & CV Splits \\ \midrule
GTZAN & 1000 & 10 & Yes & 30s & 10-fold \\
LMD & 1300 & 10 & Yes & Full & 3-fold \\ 
ISMIR & 1458 & 6 & No & Full & Train/Test \\ 
HOMBURG & 1886 & 9 & No & 10s & 10-fold \\ 
EXBALLROOM & 4180 & 13 & No & 30s & 10-fold \\ \bottomrule
\end{tabular}
\label{tab:rp_datasets}
\end{table}

Problems with the GTZAN \cite{tzan2002} dataset are well-known to the MIR community \cite{sturm2013b}. In this work, we followed the instructions proposed by Sturm \cite{sturm2013b} to minimise these problems. Such instructions include correcting mislabelled songs and maximally using the artist filter. The resulting folds are available online\footnote{\url{https://github.com/julianofoleiss/gtzan_sturm_filter_3folds_stratified.git}} to allow scientific reproducibility.

The LMD \cite{silla2008} dataset is also known to have problems regarding artist repetition and the usage of entire album in the repertoire. We used a subset of LMD that addresses these problems by applying the artist filter and an album filter, which prevents songs with the same production characteristics from being both in the training and test sets.

We did not use a cross-validation protocol with the ISMIR \cite{ismir2004} dataset because it was published with a train/test split that makes comparison to other works straightforward. We also used the HOMBURG \cite{homburg2005} dataset, which is made up of short clips and is also known for being difficult. Finally, the Extended BALLROOM \cite{marchand2016} dataset was used to test the feature sets with a larger dataset containing subsets of genres that are more similar in relation to their perceptual characteristics.

\section{Results and Discussion} 
\label{sec:rp_results}
%TODO: começar mais suave

The experimental results shown in this section allow comparing the classification performance of random features to learned and handcrafted ones. Also, they highlight the impact of changing the number of features in both the random projections and auto-encoder learning settings. Last, the results regarding transfer learning settings allow comparing random features to transferred features.

Table \ref{tab:rp_results} shows the best results obtained for all datasets and feature sets, which were consistently obtained using the SVM. All the results presented in this paper are weighed F1-Scores, that is, the weighed average of per-class F1-scores. It can be seen that, in general, MEL-RP and MEL-AE features perform better MARSYAS features for all datasets, except in EXBALLROOM.

Random projections cannot incorporate information from other domains. However, the performance improvement when comparing MEL-SPEC to MEL-RP is consistent. This means that the trade-off between the dimension reduction and the projection distortion was positive for the classification process. Results show that this trade-off could not be achieved by the PCA projection.

Both MEL-AE and MEL-RP features are not necessarily related to musical or auditory characteristics. However, since they were used in a simple, similar classification pipeline, these results reflect their frame-level descriptive capabilities from a machine learning perspective.

\begin{table}[]
\centering
\caption{Best results for every dataset (Weighted F1-Score, SVM)}
\begin{tabular}{@{}cccccc@{}}
\toprule
 & MEL-SPEC & MEL-RP & MARSYAS & MEL-AE & MEL-PCA \\ \midrule
GTZAN & 0.49 $\pm$ 0.06 & 0.62 $\pm$ 0.05 & 0.59 $\pm$ 0.05 & \textbf{0.68 $\pm$ 0.06}  & 0.22 $\pm$ 0.07 \\
LMD & 0.42 $\pm$ 0.02 & \textbf{0.77 $\pm$ 0.01} & 0.66 $\pm$ 0.03 & \textbf{0.77 $\pm$ 0.02} & 0.26 $\pm$ 0.03 \\ 
ISMIR & 0.52 & 0.81 & 0.79 & \textbf{0.86} & 0.40  \\ 
HOMBURG & 0.43 $\pm$ 0.03 & 0.49 $\pm$ 0.02 & 0.41 $\pm$ 0.02 & \textbf{0.53 $\pm$ 0.04} & 0.24 $\pm$ 0.02 \\ 
EXBALLROOM & 0.35 $\pm$ 0.03 & 0.54 $\pm$ 0.03 & \textbf{0.67 $\pm$ 0.02} & 0.55 $\pm$ 0.03 & 0.21 $\pm$ 0.02 \\ \bottomrule
\end{tabular}
\label{tab:rp_results}
\end{table}

Even though there is a clear improvement when using MEL-RP over MEL-SPEC in EXBALLROOM, the best results were achieved with MARSYAS features. This can be related to the fact that EXBALLROOM was built using subsets of tracks with a high similarity in their timbre characteristics. Within the dataset, genre is only distinguishable with respect to time-aware descriptions such as rhythm and tempo-related features. Because MEL-RP, MEL-SPEC and MEL-AE features are based solely on timbre characteristics, they are not able to describe time-dependent features.

\begin{figure}[h!]
    \centering
    \subfloat[GTZAN]{\scalebox{0.38}{\includegraphics{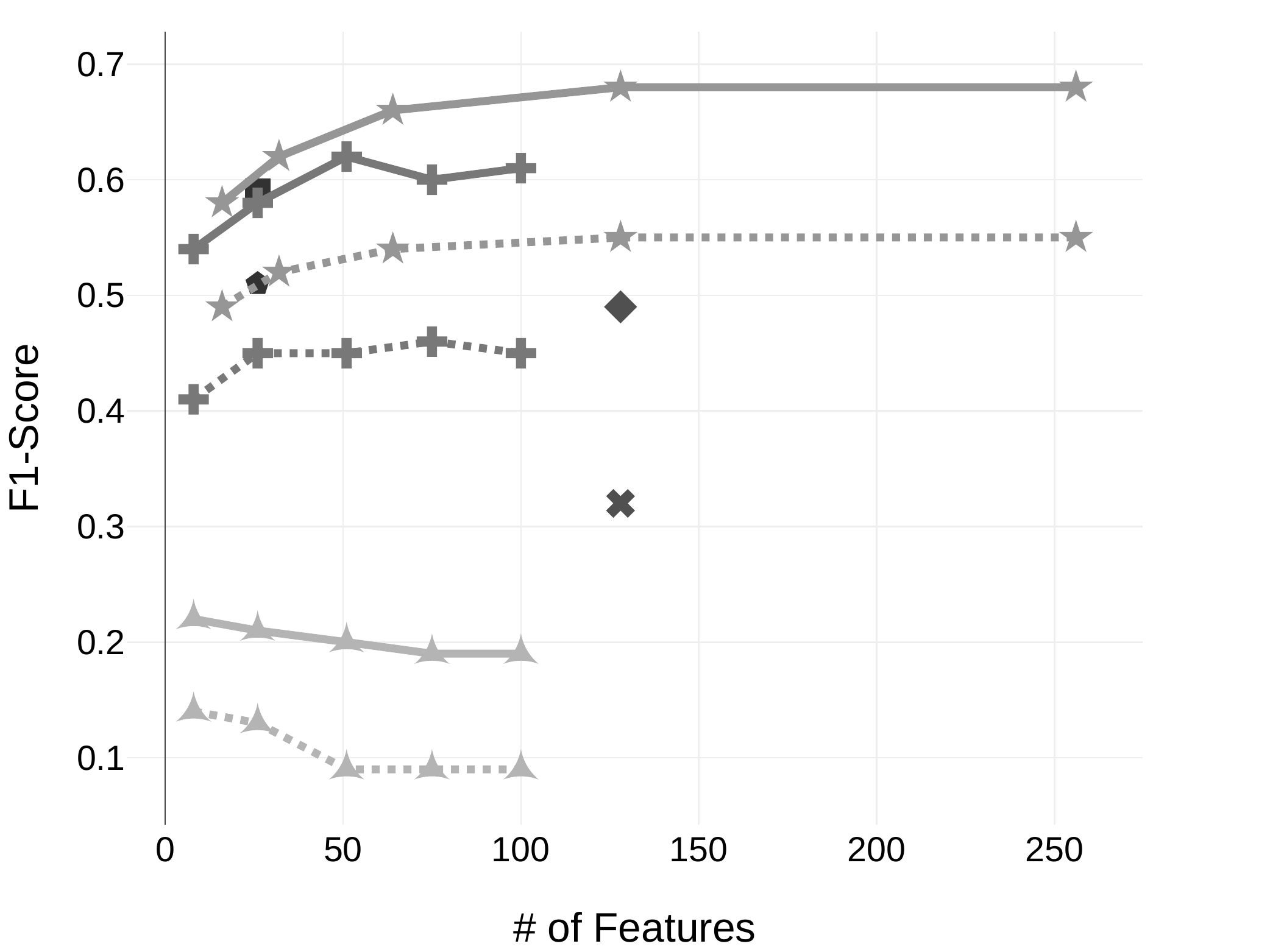}}}
    \subfloat[ISMIR]{\scalebox{0.38}{\includegraphics{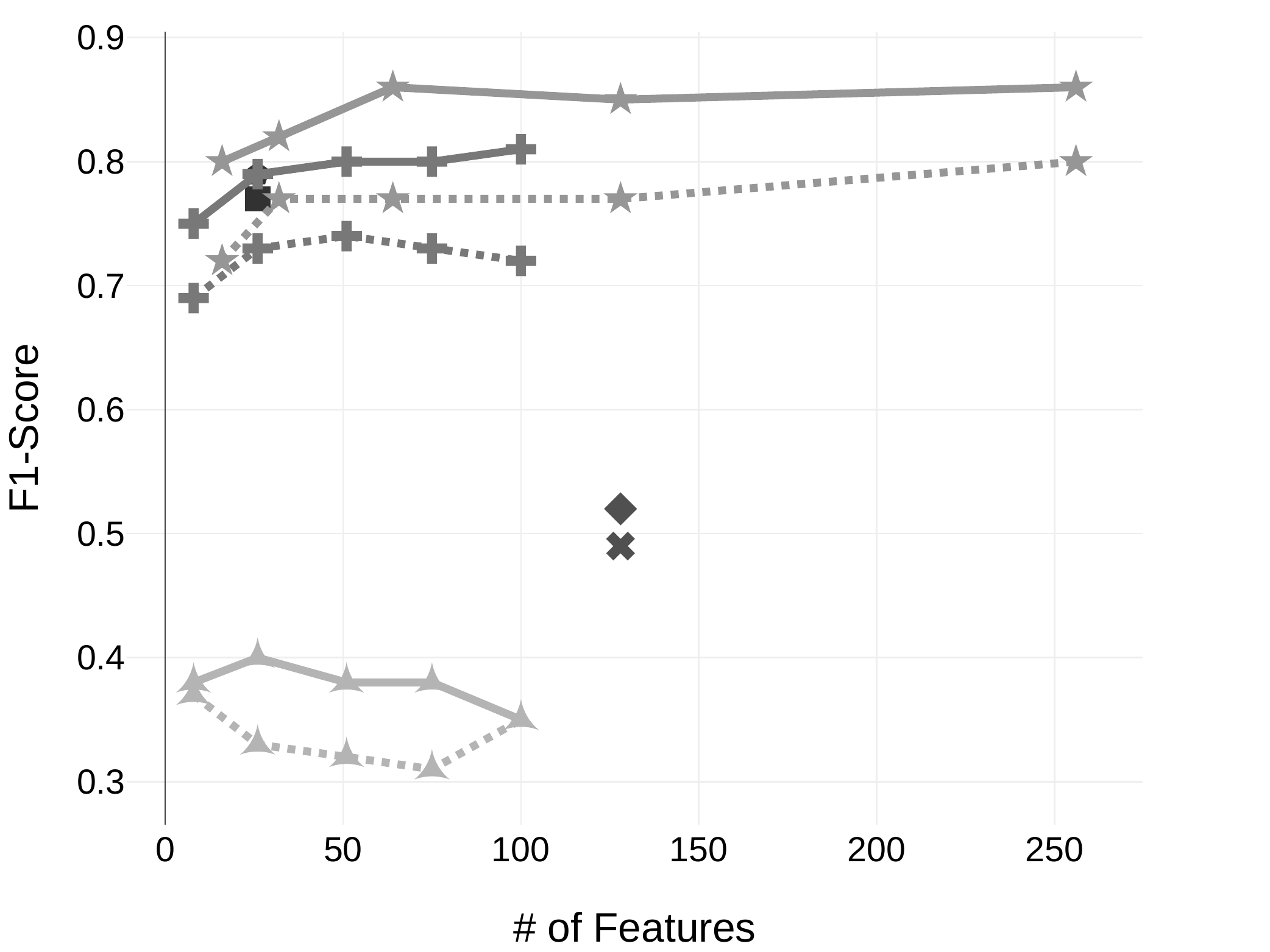}}}
    \\
    \subfloat[LMD]{\scalebox{0.38}{\includegraphics{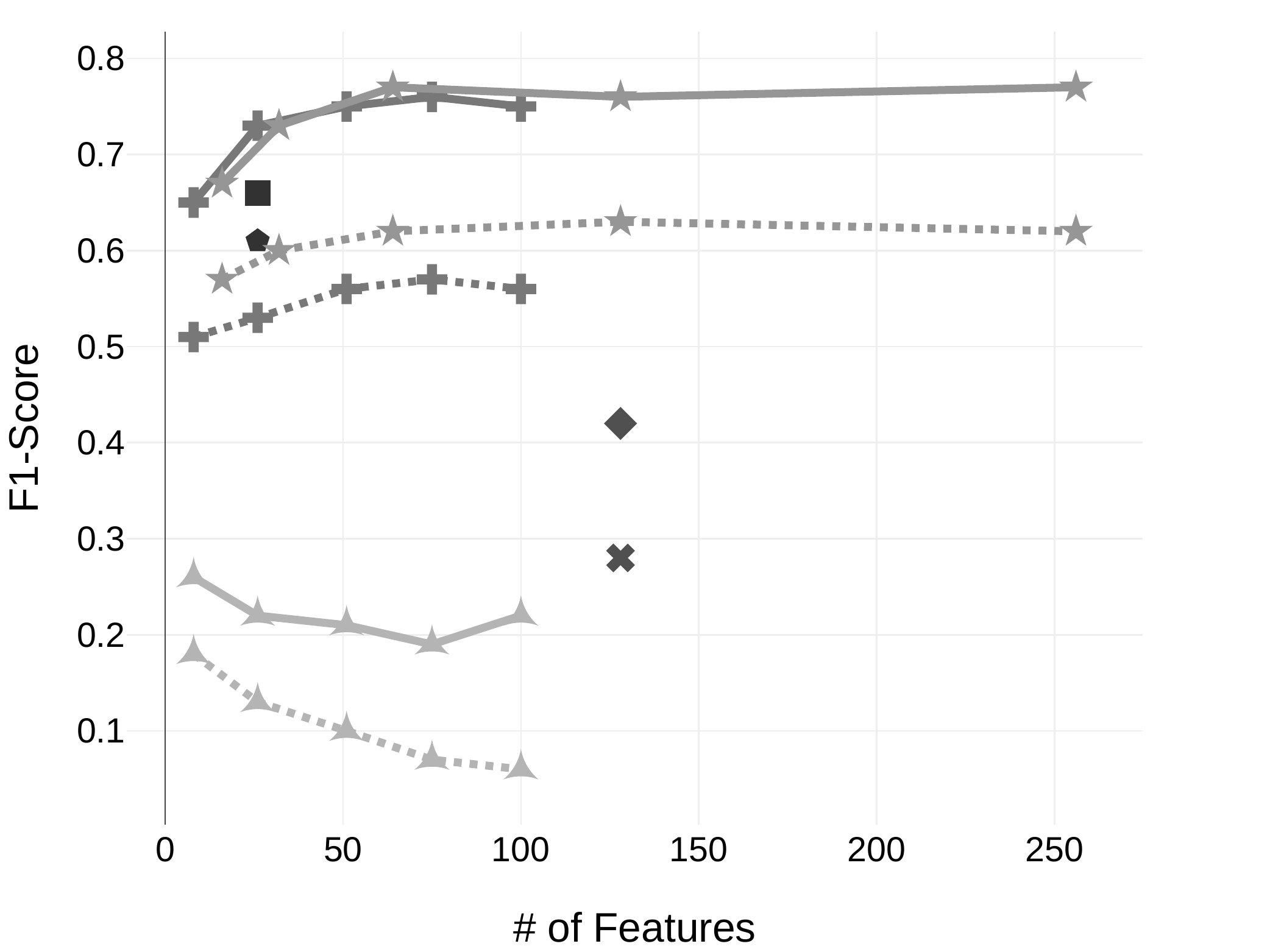}}}
    \subfloat[HOMBURG]{\scalebox{0.38}{\includegraphics{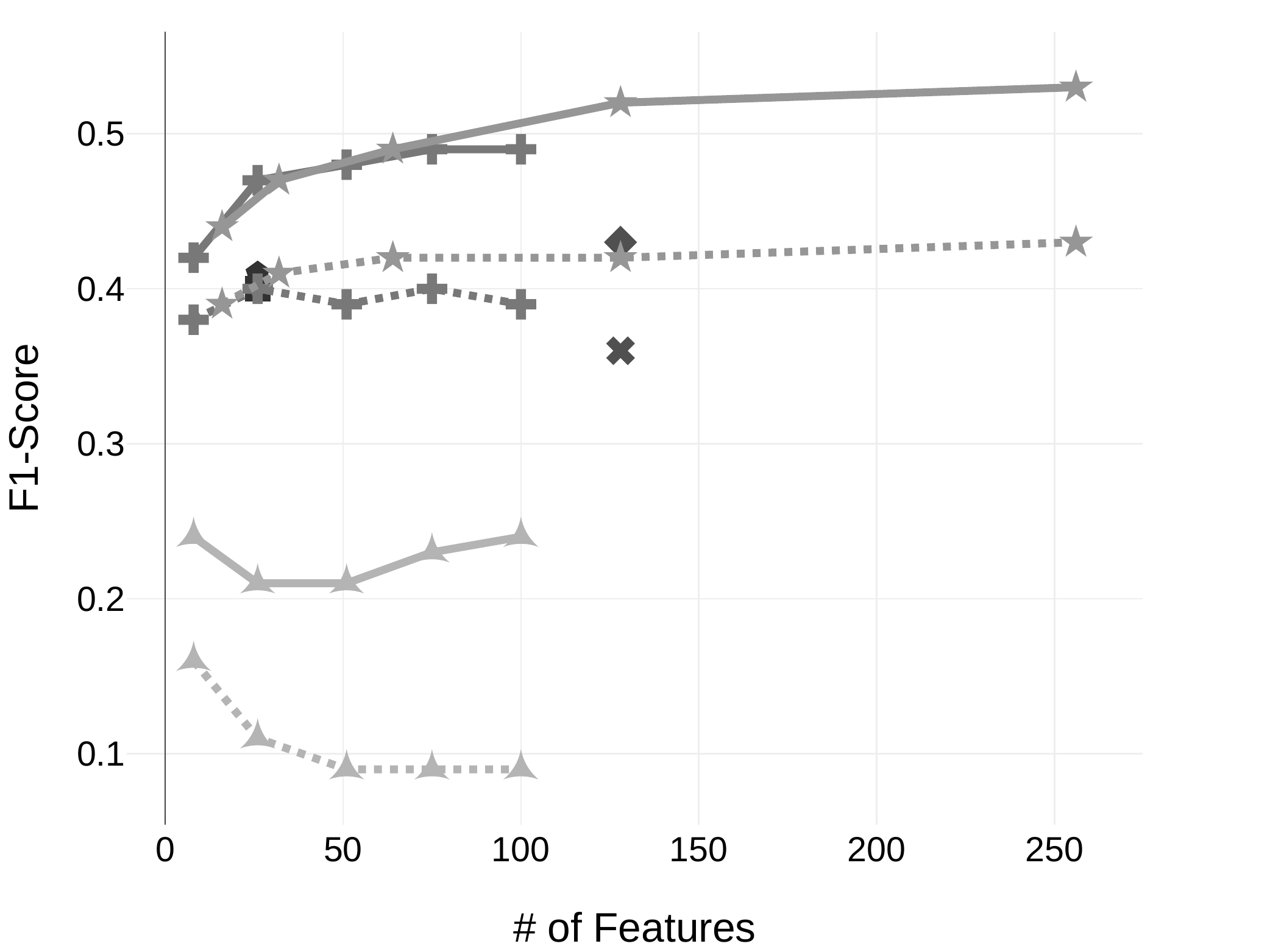}}}
    \\
    \subfloat[Extended BALLROOM]{\scalebox{0.45}{\includegraphics{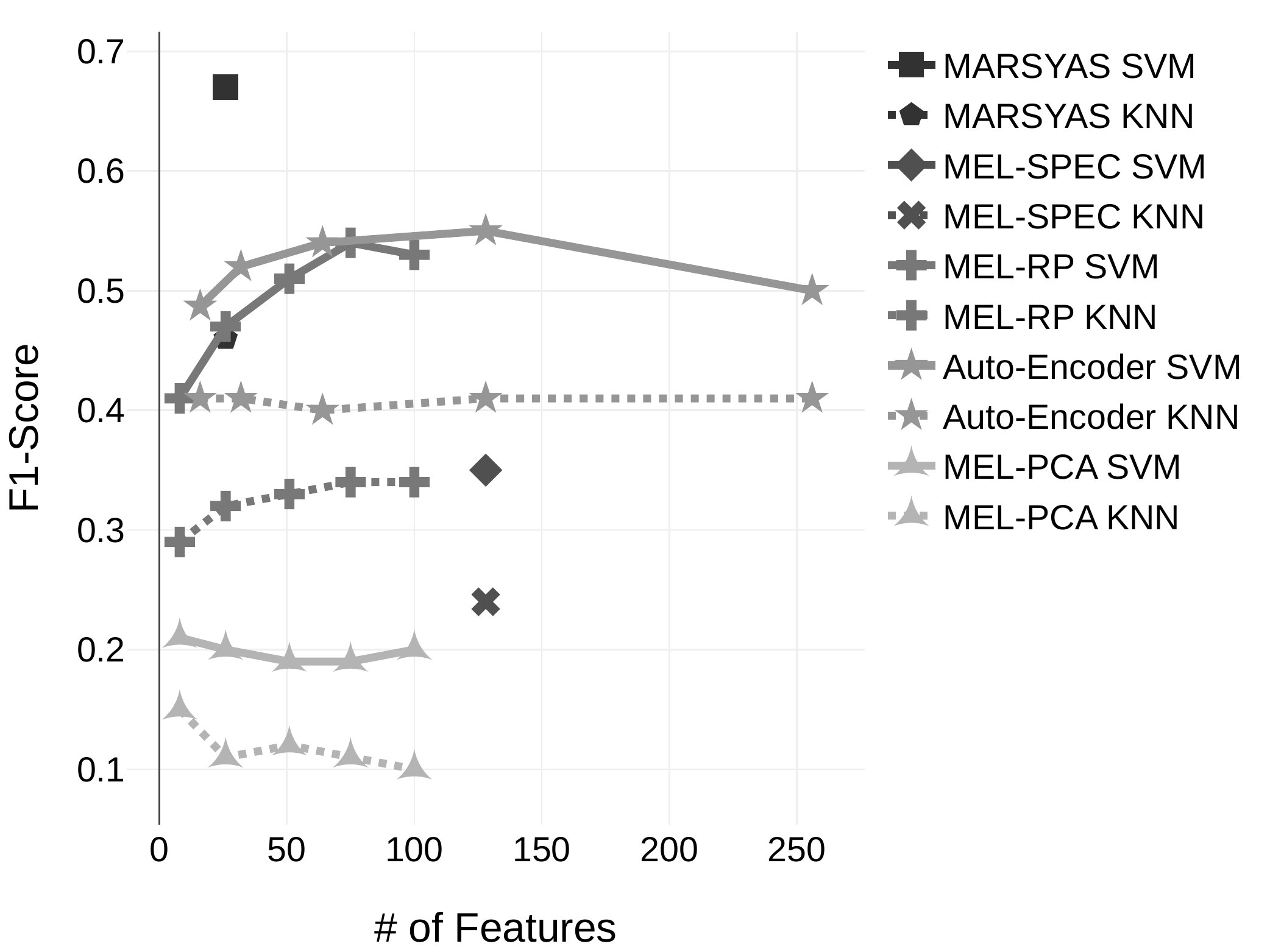}}}
    \caption{Impact of the number of features (i.e., the projection dimensionality) in the overall weighed F1-score for each dataset.}
    \label{fig:rp_features}
\end{figure}

The impact of the number of features in the classification performance was also measured, as shown in Figure \ref{fig:rp_features}. 
%The figure omits standard deviations for readability, but they are consistent to the ones shown in Table \ref{tab:rp_results}.
The number of features shown is different from the number of  features that were yielded to the classifiers, which is 12 times greater (because of the differentials, means and variances).

Figure \ref{fig:rp_features} shows that, in general, performance increases as the number of features rises for both MEL-AE and MEL-RP using SVM and KNN. However, this behaviour saturates around 50 to 100 features, leading to result saturation.

Except for Extended Ballroom, MEL-AE features achieve the best results in all feature sets. However, MEL-RP features lead to comparable results. Also, changing from SVM to KNN in the machine learning pipeline consistently shows a greater impact in the results than changing the feature set from MEL-AE to MEL-RP. Interestingly, for ISMIR and LMD, MEL-RP with KNN performs even better than MEL-SPEC with SVM, which further highlights the relevance of the random projection. 

The results regarding transfer learning settings are shown in Table \ref{tab:rp_transfer}. It can be seen that learned features lead to a significant performance drop when used in a shallow-learning scenario. This indicates that in this case features learned from a dataset are not necessarily relevant in other datasets. Also, it can be seen that the performance drop caused by using learned features is larger than the drop related to using random features (as shown in Table \ref{tab:rp_results}).

\begin{table}[h!]
\centering
\caption{Performance of learned features in different datasets (Weighed F1-Score, using SVM). Transfer learning leads to a significant performance drop.}
\begin{tabular}{@{}cccc@{}}
\toprule
Target $\downarrow$ / Source $\rightarrow$ & GTZAN & LMD & GTZAN+LMD \\ \midrule
GTZAN &\textbf{ 0.68 $\pm$ 0.06} & 0.51 $\pm$ 0.05 & 0.53 $\pm$ 0.05 \\
LMD & 0.66 $\pm$ 0.04 & \textbf{0.77 $\pm$ 0.02} & 0.66 $\pm$ 0.04 \\ \midrule
\end{tabular}
\label{tab:rp_transfer}
\end{table}

\section{Conclusion}
\label{sec:rp_conclusion}

In this work we have introduced random projection of Mel-Spectrograms (MEL-RP) as a feature set in the context of automatic music genre classification. Our results show that MEL-RP achieves results comparable to those obtained using a feature learning approach. MEL-RP, however, has the advantages of not requiring feature learning mechanisms. This reduces computing requirements during training, because the generation of a suitable random matrix is straightforward.

In perspective to handcrafted features, MEL-RP has the advantage of requiring less domain-specific knowledge. Additionally, MEL-RP outperforms MARSYAS features in most datasets. However, it leads to worse results in the EXBALLROOM dataset, whose genre classification is highly linked to rhythm properties.

Also, MEL-RP has shown to perform better than features obtained by transfer learning. This can be due to the fact that switching datasets can lean to changing the texture distribution. Such a change harms assumptions related to building an auto-encoder, but have no impact on the separation properties of random projections. This indicates that in shallow-learning systems MEL-RP is more suitable for customisation applications than using features transferred from different datasets.